\documentclass[aps,prb,twocolumn,superscriptaddress]{revtex4}
\usepackage{graphicx}

\begin{document}

\title{A local metallic state in globally insulating $La_{1.24}Sr_{1.76}Mn_2O_7$ well above the metal-insulator transition}

\author{Z. Sun}
\altaffiliation{zsun@lbl.gov}\affiliation{Department of Physics,
University of Colorado, Boulder, CO 80309, USA}\affiliation{Advanced
Light Source, Lawrence Berkeley National Laboratory, Berkeley, CA
94720, USA}
\author{J. F. Douglas}
\affiliation{Department of Physics, University of Colorado, Boulder,
CO 80309, USA}
\author{ A. V. Fedorov}
\affiliation{Advanced Light Source, Lawrence Berkeley National
Laboratory, Berkeley, CA 94720, USA}
\author{Y. -D. Chuang}
\affiliation{Advanced Light Source, Lawrence Berkeley National
Laboratory, Berkeley, CA 94720, USA}
\author{H. Zheng}
\affiliation{Materials Science Division, Argonne National
Laboratory, Argonne, IL 60439, USA}
\author{J. F. Mitchell}
\affiliation{Materials Science Division, Argonne National
Laboratory, Argonne, IL 60439, USA}
\author{D. S. Dessau}
\altaffiliation{Dessau@colorado.edu} \affiliation{Department of
Physics, University of Colorado, Boulder, CO 80309, USA}

\date{\today}

\begin{abstract}

\end{abstract}


\maketitle

\textbf{In the spectacularly successful theory of solids, the
distinction between metals, semiconductors, and insulators is based
upon the behavior of the electrons nearest the Fermi level $E_F$,
which separates the occupied from unoccupied electron energy levels.
A metal has $E_F$ in the middle of a band of electronic states,
while $E_F$ in insulators and semiconductors lies in the gap between
states. The temperature-induced transition from a metallic to an
insulating state in a solid is generally connected to a vanishing of
the low energy electronic excitations \cite{Imada}.  Here we show
the first direct evidence of a counter example, in which a
significant electronic density of states at the Fermi energy exists
in the insulating regime. In particular, angle-resolved
photoemission data from the ``colossal magnetoresistive" oxide
$La_{1.24}Sr_{1.76}Mn_2O_7$ show clear Fermi edge steps both below
the $T_C$ when the sample is globally metallic, as well as above
$T_C$ when it is globally insulating. Further, small amounts of
metallic spectral weight survive up to the temperature scale $T^*$
more than twice the $T_C$ of the system. Such behavior also may have
close ties to a variety of exotic phenomena in correlated electron
systems including in particular the pseudogap scale T* in underdoped
cuprates \cite{Timusk}.}

\begin{figure}[tbp]
\begin{center}
\includegraphics[width=1\columnwidth,angle=0]{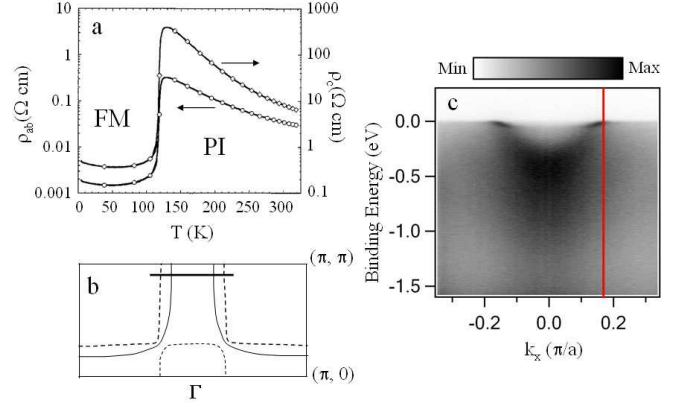}
\end{center}
\caption{Overview of features of $La_{1.24}Sr_{1.76}Mn_2O_7$. (a)
Resistivity vs. temperature, after ref 3. (b) A representative Fermi
surface. (c) Low temperature (20K) ARPES data over a
large-energy-scale taken along the black cut near the zone boundary,
as shown in (b).} \label{fig1}
\end{figure}

As shown in figure \ref{fig1}a, the colossal magnetoresistive (CMR)
oxide $La_{2-2x}Sr_{1+2x}Mn_2O_7$ (x=0.38) exhibits a metal
insulator transition at a $T_C$ of about 130K, at which point the
system also switches from being a ferromagnet (low $T$) to a
paramagnet (high $T$) \cite{LiQA}.  We performed angle-resolved
photoemission spectroscopy (ARPES) experiments on cleaved single
crystals of these materials, with an experimental arrangement as
described elsewhere \cite{Sun}. ARPES is an ideal experimental probe
of the electronic structure since it gives the momentum-resolved
single-particle excitation spectrum. As discussed in ref. 4 the
x=0.38 compound studied here does not contain the low energy
pseudogap of the x=0.4 samples \cite{DessauPRL, ChuangScience,
Saitoh, Mannella}(see supplementary material for more details on
this, the possible issue of surface sensitivity of ARPES, and of
possible intergrowths at the surface). The much larger metallic
spectral weight of these non-pseudogapped compounds also allows us
to study the electronic behavior in greater detail.

\begin{figure}[tbp]
\begin{center}
\includegraphics[width=1\columnwidth,angle=0]{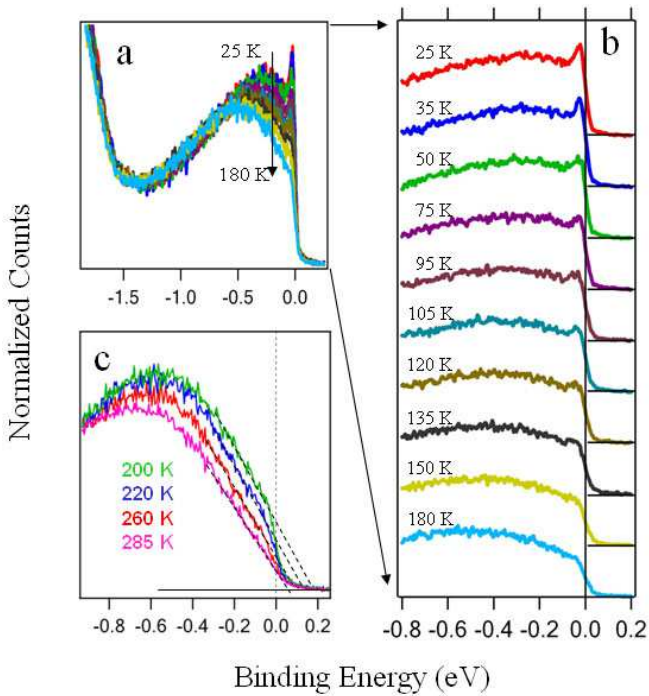}
\end{center}
\caption{Energy Distribution Curves (EDCs) as a function of
temperature at $k_F$ (red line of figure \ref{fig1}c), indicating
metallic spectral weight above $T_C$. (a, b) the same data set
scaled by the incident flux and are taken while warming. (c) EDCs
from a different sample taken at the high temperature range.  Clear
breaks are seen in the spectral intensity near $E_F$ for all but the
highest temperature, indicating finite metallic spectral weight and
a $T^*$ just above 285K (see figure \ref{fig21} for details of the
$T^*$ determination).} \label{fig2}
\end{figure}

Figure \ref{fig1}c shows a large-energy-scale experimental picture
of a low temperature $d_{x^2-y^2}$ symmetry band taken along the
black cut near the zone boundary, as shown in figure \ref{fig1}b. We
are able to get clean data by isolating the various bilayer-split
bands using different photon energies, as described in ref. 4. In
particular, in this paper we only show data from the antibonding
bilayer-split band which has Fermi crossings at $k_x$=$\pm$0.17
$\pi/a$, $k_x$=0.9 $\pi/a$, corresponding to the solid Fermi surface
in figure \ref{fig1}b. The energy distribution curves (EDCs) at
$k_F$ (indicated by the red line in figure \ref{fig1}c) taken at a
series of temperatures are shown in figure \ref{fig2}a. Figure
\ref{fig2}b shows the identical spectra and identical scaling, but
offset vertically for clarity. All spectra have been normalized only
to the incident photon flux.

At low temperature, the EDCs clearly show a structure of
peak-dip-hump, where the peak and the hump would nominally be
considered the coherent part (quasiparticle) and ``incoherent" part
of the single particle spectrum respectively, as has been discussed
for the spectra of the high $T_C$ cuprate superconductors
\cite{Dessau, ShenDessau, Damascelli}. One sees that the near-$E_F$
spectral weight diminishes with increasing temperature, while the
high binding energy ($>$700meV) part is less affected by
temperature. Contrary to the general picture of the metal-insulator
transition, in which a gap develops in the single particle spectrum
when an electronic system becomes insulating \cite{Imada}, the EDCs
here still exhibit a sharp Fermi cutoff indicating metallic behavior
at temperatures in which the macroscopic DC conductivity is
characteristic of insulation (e.g. the spectra at 135, 150 and
180K). To our knowledge, this unusual behavior, a metallic Fermi
edge in a globally insulating system, has not been previously
observed on the insulating side of a metal-insulator transition. The
opposite, in which a metallic system shows a lack of a Fermi cutoff,
is on the other hand expected in exotic low-dimensional systems such
as the Luttinger Liquids \cite{Voit}, and has likely been observed
\cite{Allen}.  The other situation most likely to show a metallic
Fermi edge in a globally insulating system is that of an
Anderson-localized system beyond the mobility edge. However, in such
systems a Coulomb gap is expected to remove the metallic weight near
the Fermi energy \cite{Varma}. Our data could only be consistent
with such a scenario if the Coulomb gap were extremely small - on
the order of a few meV or less.  Such a picture also would not
naturally explain the metallic spectral weight dependence with
temperature, to be discussed in more detail later in the letter.

\begin{figure}[tbp]
\begin{center}
\includegraphics[width=0.7\columnwidth,angle=0]{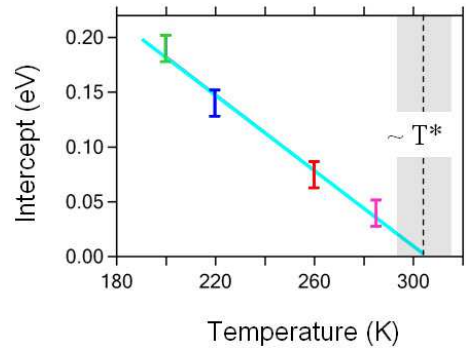}
\end{center}
\caption{Determination of $T^*$.  Zero intensity intercepts as a
function of temperature from the linear fits to the data shown in
figure \ref{fig2}c.  These intercepts go to zero at 305K, which we
label as $T^*$ $\sim$ 305K.} \label{fig21}
\end{figure}

On a different sample we have done higher temperature scans, looking
for a possible temperature scale at which the metallic spectral
weight disappears. These data are shown in figure \ref{fig2}c and
show a clear discontinuity in the slope near the Fermi energy for
all but the 285K data, indicating a finite metallic spectral weight.
This effect is emphasized by an extrapolation of the spectral weight
using a simple linear fit to the data between -0.3 and -0.05 eV, as
shown by the dotted lines in the figure. Upon raising the sample
temperature we see that the intercept of these dotted lines with the
horizontal axis decreases at an approximately linear rate (figure
\ref{fig21}).  As shown in this figure these zero intensity
intercepts reach the Fermi energy at 305 $\pm$10K. We thus indicate
305K as the temperature at which the first bits of metallic weight
become apparent, which we indicate as the temperature $T^*$.
Technical reasons including sample aging and excessive manipulator
drift preclude us from making the full range of measurements on a
single cleave. We therefore used different samples to study the
electronic excitations in different temperature regimes.

\begin{figure}[tbp]
\begin{center}
\includegraphics[width=1\columnwidth,angle=0]{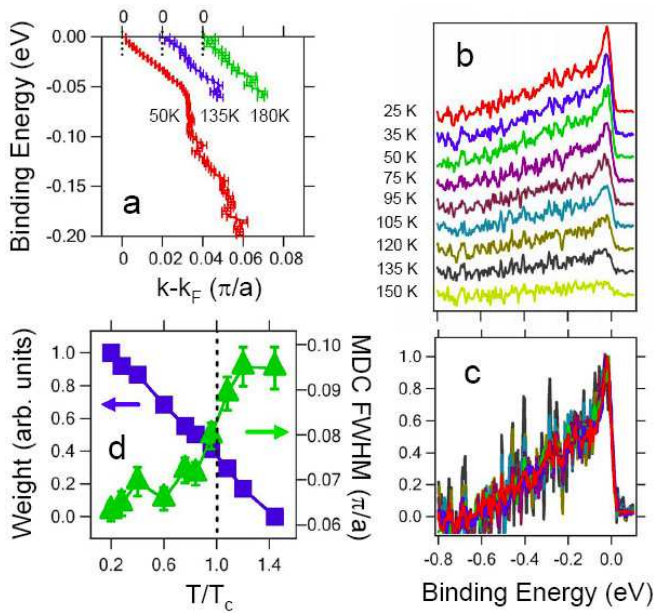}
\end{center}
\caption{Properties of the metallic portion of the sample.  (a)
Electronic dispersion showing a similar $k_F$, $v_F$ and
electron-phonon coupling   as a function of temperature.  (b, c)
Metallic EDCs or M-EDCs obtained by subtracting the 180K EDC from
all lower temperature data. (b) shows the raw scaling while (c)
scales each spectrum to have a similar max intensity. (d) MDC widths
(green triangles) and integrated M-EDC spectral weights (blue
squares) as a function of temperature.} \label{fig3}
\end{figure}

Figure \ref{fig3}a shows the electronic dispersion of the near-Fermi
states as a function of temperature obtained from an analysis of
momentum distribution curves (MDCs).  This data indicates that the
main properties of the metal, such as the Fermi wave vector $k_F$,
the Fermi velocity $v_F$, the electron phonon coupling parameter
$\lambda$, and the effective mass $m^*$ don't change significantly
as a function of temperature, even as the metal-insulator transition
temperature $T_C$ is traversed.  This is an unexpected behaviour for
a metal-insulator transition in which these parameters would vary
dramatically with temperature, and likely even diverge \cite{Imada}.

Our data can be understood by invoking a model of disconnected local
ferromagnetic metallic regimes above  $T_C$ up to approximately the
temperature $T^*$. This suggestion is consistent with earlier
studies which have found significant ferromagnetic signals far above
$T_C$ \cite{ArgyriouJAP,Osborn,Rosenkranz1,Rosenkranz2}, since
metallicity and ferromagnetism should have a connection in these
systems via the double-exchange interaction. In general, the
metallic regions may be either phase separated (and possibly static)
domains \cite{Uehara,Dagotto} or they may be dynamic fluctuations of
the ferromagnetic metallic state, which in a two-dimensional system
may persist to quite high temperatures
\cite{ArgyriouJAP,Osborn,Rosenkranz1,Rosenkranz2}. We will discuss
these two possibilities later in this letter. Here we show that we
can study the metallic portions further by our ability to
approximately deconvolve the spectrum into the components which
arise from the metal and non-metal portions. We do this by
subtracting the 180K EDC from all other EDCs as shown in figure
\ref{fig2}b to create ``metallic EDCs" or M-EDCs as shown in figure
\ref{fig3}b. It should be pointed out that the slight variation of
spectra from sample to sample, which has been commonly observed in
ARPES, imperils the practice of extracting data of one sample from
that of another. Therefore, we don't use the higher temperature data
of figure \ref{fig2}c to do the subtraction as this is from a
different sample. Figure \ref{fig3}c shows the same M-EDCs but
scaled to all have the same amplitude. Within the noise, all the
M-EDCs have similar lineshapes with coherent peaks near $E_F$ and an
incoherent background at high binding energy, though the widths of
the M-EDC coherent peak (or low energy MDC peak) become broader with
increasing temperature (figure \ref{fig3}d). The integrated spectral
weight of the M-EDCs varies smoothly as a function of temperature,
with no clear break at $T_C$ (figure \ref{fig3}d). This, as well as
the approximate temperature-independence of the M-EDC lineshape
indicates that the electrons in the metallic regions have similar
properties above and below $T_C$, and that temperature has
surprisingly little effect on the behavior or interactions of
electrons in the metallic regions. This is consistent with the
approximate independence of $v_F$, $\lambda$, and $m^*$ in the
metallic regions shown in figure \ref{fig3}a. The experimentally
determined MDC width of the electrons at the Fermi energy (green
triangles of figure \ref{fig3}d) does broaden with increasing
temperature. The inverse of this quantity, the mean free path of the
electrons, thus decreases with increasing temperature, consistent
with a decreased size of metallic regions or increased scattering
events at higher temperatures.

While many aspects of our data are consistent with either the phase
separation or magnetic fluctuation picture, certain aspects of it
can address the question of whether the metallic regions above $T_C$
are phase-separated out from a more insulating environment
\cite{Uehara,DagottoBook}, or if they are just fluctuations from a
lower temperature ordered environment which is otherwise homogeneous
\cite{Rosenkranz1,Rosenkranz2}. In particular, the smooth dependence
of the spectral weight of the metallic regions as a function of
temperature across $T_C$ (blue squares of figure \ref{fig3}d) is
more consistent with phase separation, as we would likely expect a
clear drop in the metallic weight near $T_C$ if the metallic
portions were just fluctuations of the ordered lower temperature
environment. At other doping levels (for example $x$=0.4),
experiments do observe a sharp drop in the metallic weight at $T_C$
to zero or almost zero \cite{Mannella}, and so those samples may be
more consistent with the fluctuation physics.

Within the picture of phase separation, we imagine that the metallic
islands arise at a temperature $T^*$ near room temperature, which
also may be related to the temperature scale at which polaronic
correlations freeze \cite{ArgyriouPRL}.  As the temperature is
lowered the size and proportion of metallic portions grows until a
critical ratio of metallic to insulating portions is reached.  At
that point electrons can percolate from one metallic region to
another, bringing about the macroscopic metallic \cite{Uehara} and
ferromagnetic states, as well as being consistent with the
``colossal" decrease in resistivity with an applied magnetic field.
In certain models this behavior is expected from a competition
between different phases, for example between the ferromagnetic
metal phase and the charge-ordered antiferromagnetic insulating
phase \cite{Uehara, Dagotto, Tokura}, though in contrast to ref 19,
the materials used here are far away from the charge-ordered doping
level. Theoretical arguments predict both the phase separation and
the existence of a higher temperature scale $T^*$  \cite{Burgy},
with ideas similar to the Griffiths singularity \cite{Griffiths}  in
which $T^*$ would be the critical temperature of the associated
clean system in the absence of disorder, and which have recently
been discussed in the context of manganite physics \cite{Burgy,
Salamon}. We are presently undertaking a more thorough study of the
full Fermi surface to test this percolation model quantitatively.

A $T^*$ scale is one of the key properties of the high $T_C$
superconductors, and has for years been the subject of intense
controversy \cite{Timusk}. In these compounds, disorder also appears
to be highly relevant, especially in the underdoped regime where the
$T^*$ scale exists. In that case it signals the emergence of the
pseudogap, which may be the precursor to the long range
superconducting order which forms at $T_C$ \cite{Emery} -- a clear
analogy to the manganites where $T^*$ signals the emergence of the
metallic domains which become long range at $T_C$. Also similar to
the cuprates, it appears that the $T^*$ temperature scales may not
be universal to all doping levels of the manganites. Pinning these
details down and then understanding their implications will
certainly be an area of intense study in the near future.

It is becoming increasingly clear that some of the most dramatic
responses in modern materials occur in systems in which multiple
phases or orders with similar energy scales compete with each other
\cite{Dagotto, Tokura, Burgy, Murakami}.  It is then natural that in
at least some of these systems spatial heterogeneities will occur,
and small perturbations can cause drastic macroscopic alterations to
the physical properties or even new types of  ``emergent" behavior.
The key is finding which aspects of the inhomogeneity are intrinsic
and what is their role in determining the key physical properties of
the system.


The authors thank Y. Tokura and T. Kimura for providing preliminary
samples and are grateful to D. N. Argyriou, A. Bansil, E. Dagotto,
K. Gray, A. Moreo, R. Osborn, L. Radzihovsky, D. Reznik, S.
Rosenkranz, Y. Tokura, and M. Veillette for helpful discussions.
This work was supported by the U.S. Department of Energy under grant
DE-FG02-03ER46066 and by the U.S. National Science Foundation grant
DMR 0402814. The ALS is operated by the Department of Energy, Office
of Basic Energy Sciences. Argonne National Laboratory, a U.S.
Department of Energy Office of Science Laboratory, is operated under
Contract No. DE-AC02-06CH11357. The U.S. Government retains for
itself, and others acting on its behalf, a paid-up nonexclusive,
irrevocable worldwide license in said article to reproduce, prepare
derivative works, distribute copies to the public, and perform
publicly and display publicly, by or on behalf of the Government.

\textbf{Supplementary discussion:}

1. \emph{The difference between $x$=0.38 and $x$=0.4 samples.} It
should be pointed out that there is a remarkable difference between
the ARPES spectra of $La_{1.24}Sr_{1.76}Mn_2O_7$ and
$La_{1.2}Sr_{1.8}Mn_2O_7$ samples, even though many macroscopic
properties are similar. Quasiparticles have been found near the zone
boundary at the doping levels of $x$=0.36 and 0.38 in
$La_{2-2x}Sr_{1+2x}Mn_2O_7$, while there exists a large energy
pseudogap in $x$=0.40 samples
\cite{Sun,DessauPRL,ChuangScience,Saitoh,Mannella}. Temperature
dependent studies have also been performed on $x$=0.4 samples and
have not shown evidence for metallic spectral weight above $T_C$
\cite{ChuangScience,Saitoh,Mannella}. Similar to high-$T_C$
cuprates, physical properties exhibit strong variations with doping
in manganites. The cause of the difference between
$La_{2-2x}Sr_{1+2x}Mn_2O_7$ ($x$=0.38) and
$La_{2-2x}Sr_{1+2x}Mn_2O_7$ ($x$=0.40) samples is not understood
yet, though it could have to do with the increased lattice anomalies
for the 0.4 samples [J. Mitchell et al., J. Phys. Chem. B 105, 10731
(2001).], the onset of spin canting between ferromagnetic layers
which starts at the doping level of 0.4 [M.Kubota et.al., J. Phys.
Soc. Jpn 69, 1606 (2000).],  or even something extrinsic such as a
surface issue.

2. \emph{The issue of surface sensitivity.} Because of the shallow
probing depth of the ARPES experiment ($\sim$ 5-10 Angstroms), we
cannot completely rule out the potential that a surface phase whose
properties do not follow those of the bulk gives rise to some of the
phenomena reported here. For ARPES on the layered manganites we are
relatively well off since the samples cleave readily between the
$La,Sr-O$ bilayers, which are ionically (not covalently) bonded.
High quality LEED pictures without any evidence of surface
reconstruction are obtained from these surfaces. The doping level at
the surfaces, as obtained from the Fermi surface volume also appears
to be correct for these samples - for example the $d_{x^2-y^2}$
bonding band Fermi surface nesting vector of $0.27\times(2\pi/a)$
for the $x=0.38$ samples used in this study \cite{Sun} exactly
matches that obtained from neutron scattering measurements
\cite{ArgyriouPRL}. The nesting vector of 0.4 samples is slightly
larger at $0.3\times(2\pi/a)$ \cite{ChuangScience} and also matches
the results of scattering measurements [L. Vasiliu-Doloc, et al.
Phys. Rev. Lett. 83, 4393 (1999).].

3. \emph{The issue of intergrowths.} One should consider whether it
might be possible for the metallic spectral weight far above $T_C$
to have originated from small bits of intergrowth (IG) of a higher
$T_C$ sample left near the surface after cleaving.  Here we discuss
why that is inconsistent with our data. Such intergrowths should not
arise from a layered manganite, as the maximum temperature at which
bulk metallic behavior is found among all known layered manganites
is $\sim$ 160K. A small amount of (non-layered) perovskite-like IG
with a $T_C$ = 300K could exist at a cleaved surface, though would
not show the bilayer splitting since the perovskite samples have
only one $MnO_2$ plane per unit cell. Both our high and low
temperature data display this bilayer band splitting (this paper
only presents the data from the antibonding component), which is a
direct consequence of having two $MnO_2$ planes per unit cell. In
addition to being able to vary the intensity of the bilayer split
bands (relative and overall) by taking advantage of the
photoemission matrix elements, we can follow the dispersion in $E$
and $k$ of each of the bilayer bands, including tracking them all
the way to $E_F$. Therefore we know the origin of the metallic
weight as explicitly originating from these bilayer split bands, and
therefore, from the bilayer manganite.

\end{document}